\documentclass[aps,apl,a4paper,twocolumn,superscriptaddress,epsf]{revtex4}
\usepackage{graphicx}
\usepackage{dcolumn}
\usepackage{bm}
\usepackage{amsmath}

\begin{document}

\hyphenation{Schwer-punkt-pro-gramm}


\title{Atom Chips: Fabrication and Thermal Properties}


\author{S. Groth}
\author{P. Kr\"uger}
\author{S. Wildermuth}
\author{R. Folman}
\author{T. Fernholz}
\affiliation{Physikalisches Institut, Universit\"at Heidelberg,
69120 Heidelberg, Germany}
\author{D. Mahalu}
\author{I. Bar-Joseph}
\affiliation{Department of Condensed Matter Physics, Weizmann
Institute of Science, Rehovot 76100, Israel}
\author{J. Schmiedmayer}
\affiliation{Physikalisches Institut, Universit\"at Heidelberg,
69120 Heidelberg, Germany}

\date{\today}

\begin{abstract}
Neutral atoms can be trapped and manipulated with surface mounted
microscopic current carrying and charged structures. We present a
lithographic fabrication process for such atom chips based on
evaporated metal films. The size limit of this process is below
1$\mu$m. At room temperature, thin wires can carry more than
10$^7$A/cm$^2$ current density and voltages of more than 500V.
Extensive test measurements for different substrates and metal
thicknesses (up to 5 $\mu$m) are compared to models for the
heating characteristics of the microscopic wires. Among the
materials tested, we find that Si is the best suited substrate for
atom chips.

\end{abstract}


\maketitle



Manipulation of neutral atoms using micro-structured surfaces
(atom chips) has attracted much attention in recent years
\cite{Fol02}. Atom chips promise to combine the merits of
microfabrication and integration technologies with the power of
atomic physics and quantum optics for robust manipulation of
atomic quantum systems. Extreme and precise confinement in traps
with large level spacing is possible. Neutral atoms have been
trapped by currents flowing in microscopic wires fabricated on the
atom chip surface \cite{Rei99,Fol00}, manipulated by electric
fields \cite{Kru03}, and even cooled to Bose-Einstein condensates
\cite{Ott01,Hae01b,Sch03}.

The demands on atom chips are high current densities to create
steep traps, and small structure sizes to create complex
potentials at a scale where tunneling and coupling between traps
can become important. Exceptional high-quality fabrication is
essential, since the smallest inhomogeneities in the bulk of the
wire or the fabricated edges can lead to uncontrolled deviations
of the current flow and therefore to disorder potentials in the
magnetic guides and traps \cite{For02,Jon03,Lea03,Wan04,Est04}.

The fabrication process preferred by groups working in the field
is to grow thick wires from a thin patterned layer using
electroplating \cite{Drn98,For02b,Lev03}, which allow large
currents for trapping and manipulating the atoms.

\begin{figure}
\includegraphics[angle=0,width=\columnwidth]{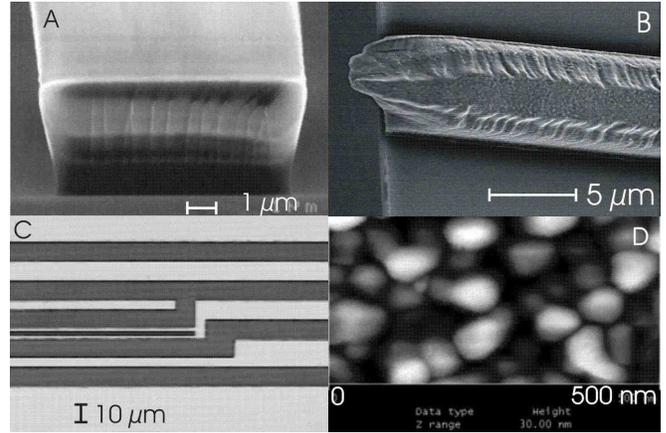}
{\caption{Microscope images of chip details during and after the
fabrication. A: SEM picture of the resist structure. Its thickness
is $4.5\mu$m, the undercut is $0.6\mu$m. B: SEM image of a typical
fabricated wire. C: 1, 5, and $10 \mu$m wide Gold wires on a fully
fabricated chip. D: AFM picture of the gold surface. The grain
size is 50--80nm.}\label{preparepic}}
\end{figure}

In contrast, we have chosen to directly pattern up to 5$\mu$m tall
evaporated high-quality layers of gold using a photolithographic
lift-off technique (Fig.\ \ref{preparepic}).
Wires are defined by thin gaps in the evaporated gold surface.
These $\mu$m sized gaps only produce an insignificant amount of
stray light when the gold surface is used as a (nearly perfect)
mirror for laser cooling and atom imaging. This technique was
preferred because it results in high surface quality and very
smooth structure edges.

In our process, the substrate (Si or GaAs covered with a SiO$_2$
insulation layer or sapphire) is prepared with an adhesion primer.
To allow the evaporation of thick metal layers, we spin up to
5$\mu$m thick films of image reversal photoresist (AZ 5214E) onto
the sample at low speed. The resist is then exposed to UV-light
through an e-beam patterned mask. After developing the resist
structure (Fig.\ \ref{preparepic}A), a  Ti adhesion layer (35nm)
and a thick Au layer (1--5$\mu$m) are evaporated at a short
distance from the source at a rate of 5--40\AA/s. To achieve good
surface quality, care has to be taken in controlling the
evaporation speed and substrate temperature. The gold covered
resist structure is then removed in a lift-off procedure using
acetone (if necessary in a warm ultrasonic bath) and isopropanol
as solvents. A second gold layer can be added by repeating the
above process. Some chips were covered with a thin protective
insulation layer of Si$_3$N$_4$. Finally, the chips are cut or
cleaved to the desired dimensions of $25\times 30$mm$^2$.

The resulting gold surfaces (Fig.\ \ref{preparepic}D) are smooth
(grain sizes $<80$nm), and the wire edges are clearly defined. The
surface quality depends on adhesion properties and the substrate
smoothness. Semiconductor substrates (Si and GaAs) gave better
results than sapphire samples.

An important characteristic of atom chips is how much current and
voltage the microscopic wires can carry. The achievable
confinement of the atoms depends on the maximal potential
gradient, which scales with the achievable current density $j$.

\begin{figure}
\includegraphics[width=\columnwidth]{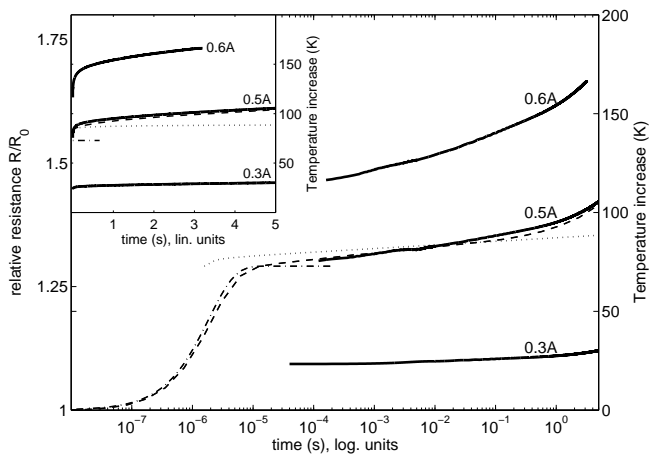}
{\caption{Temperature evolution of a
  5$\mu$m wide, 1.4$\mu$m tall wire mounted on a 700$\mu$m thick
  Si substrate with a 500nm SiO$_2$ isolation layer. The
  solid curves show measured data for 0.6A, 0.5A, and
  0.3A current pulses. In one case (0.5A) also the theoretical
  predictions (without fitting parameters) are shown. The initial
  fast temperature increase (thin dashed-dotted curve)
  occurs on a $\mu$s timescale. The analytical model
  for the heat transport through the substrate (dotted curve)
  holds only as long as the approximation of a half space
  substrate is valid. A two-dimensional
  numerical model (dashed curve) accurately reproduces the measurements.}
  \label{datapic}}
\end{figure}

To evaluate the fabrication process and to collect comprehensive
data for the operation of the atom chips and their limits, a
series of test chips was built incorporating 2mm long wires with
widths of 2, 5, 10, 50, and 100$\mu$m and heights ranging from
1-5$\mu$m on different substrates (Si with 20nm or 500nm thick
isolation layer, GaAs, and sapphire). The test chips were mounted
on a simple sample holder, and the current and voltage
characteristics were measured under moderate vacuum conditions
($10^{-6}$mbar).

Regarding charges, both semiconducting (Si) and insulating
(sapphire) substrates tolerated voltages of $>300$V ($>500$V for
sapphire) across a gap of 10$\mu$m. This provides ample
flexibility for manipulation of atoms in comparably deep
potentials even at relatively large distances from the surface by
means of electrostatic fields \cite{Kru03}.

The current characteristics were measured by pushing a constant
current through the wire and recording the voltage drop, which
yields the resistance of the wire. The experiments were carried
out in a pulsed manner similar to the real atom chip experiments,
allowing a cool down time (typically $\sim 10$s) between the
pulses. Usually, we arranged the length of the current pulses such
that the resistance rose by less than 50\%. This proved to be a
safe procedure without damaging the wires. At stronger heating,
the measurements were partly irreproducible (sometimes the wires
were even destroyed).

The measurements show two timescales for the heating process
(Fig.\ \ref{datapic}). Immediately after switching the current,
the wires heat up on a $\mu$s timescale, seen as an increased
resistance in our experiments. This initial jump, which was not
resolved in our measurements, is followed by a slow rise of
temperature, which was observed over the full duration of the
current pulses (up to 10s).




\begin{figure}
    \includegraphics[width=\columnwidth]{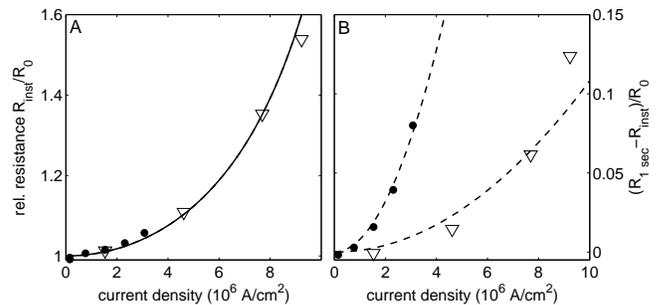}
    {\caption{Comparison of the measured heating of wires of different
    widths [$5\mu$m ($\bigtriangledown$), $50\mu$m ($\bullet$)]
    to the two models (solid and dashed lines; no
    fitting parameters used). A: The fast heating process depends
    solely on the current density. B: For the slow process, the wider
    wire exhibits a larger temperature increase at equal current
    density.}\label{crosspic}}
\end{figure}

In order to gain insight into the relevant processes and
parameters, we compared our data with a simple model
for the heating of the substrate mounted wires. The heat created
in a wire of height $h$ and width $w$ carrying a current density
$j=I/wh$ is given by ohmic dissipation. Essentially, the heat is
removed through heat conduction to the substrate, as thermal
radiation is negligible for the observed temperatures.
The temperature evolution of the wire is determined by the heat
flow through the interface, which exhibits a thermal contact
resistance (thermal conductance $k$), and the heat dissipation
within the substrate, governed by the heat conductivity $\lambda$
and heat capacity (per volume) $C$.
This leads to two very different timescales for the heat removal:

The timescale of the first process (heat flow from the wire to the
substrate through the isolation layer) is given by
$\tau_\mathrm{fast}=\frac{C_Wh}{k-hj^2\alpha\rho}$ where $C_W$ is
the heat capacity (per volume) of the wire, $\rho$ its (cold)
resistivity with a linearly approximated temperature coefficient
$\alpha$. For typical parameters of our chips, this timescale
($\sim 1\mu$s) is so short that the temperature difference between
the wire and the substrate
\begin{equation}
\Delta T_f(t)=\frac{h\rho j^2}{k-hj^2\alpha\rho}
(1-e^{-t/\tau_\mathrm{fast}})
\end{equation}
saturates practically instantaneously, unless $j$ exceeds the
limit of $\sqrt{k/h\alpha\rho}$. In this case, an exponential rise
of the temperature will lead to an almost instantaneous
destruction of the wire.

\begin{figure}
    \includegraphics[angle=0,width=\columnwidth]{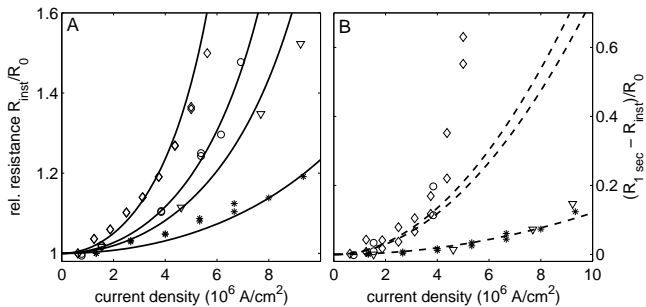}
    {\caption{Heating of $5\mu$m wide Au wires fabricated on a commercial
    GaAs wafer
    ($\diamond$), sapphire without isolation layer ($\circ$), and Si substrates with a 500nm
    ($\bigtriangledown$) and a 20nm ($\ast$) thick isolation
    layers. The heights of the wires are
    1.4 $\mu$m (Si), 2.6 $\mu$m (sapphire) and 3.2 $\mu$m (GaAs).
    A: The prediction of the model
    for the fast process (solid lines) is compared to the wire resistances (in
    units of their respective cold resistance $R_0$)
    measured a few ms after the beginning of a current pulse. The
    thermal contact resistance ($k\sim10^6$W/Km$^2$) was used as a fitting parameter.
    B: Slow heating process data taken after 1s of current
    flow. No fitting parameters were used to compare the model (dashed lines).}
    \label{substrpic}}
\end{figure}

Our model for the fast heating process quantitatively agrees with
the data. While the initial temperature rise is independent of the
wire width (Fig.\ \ref{crosspic}A), it depends on the contact
resistance to the substrate (Fig.\ \ref{substrpic}A). The latter
effect can clearly be seen by comparing the data for two Si
substrates with different insulation layers. A 500nm thick SiO$_2$
layer leads to stronger heating than a 20nm thick layer. In order
to find the optimal substrate, the values for the thermal
conductance $k$ were obtained from fitting the model to our data
($k=6.5, 3.5, 2.6, 2.3 \times 10^6$W/Km$^2$ for Si with 20nm
SiO$_2$, sapphire, Si with 500nm SiO$_2$, and GaAs, respectively).
The fitting is necessary, because $k$ includes the conductance
through the insulation layer and the contact resistances at the
individual material interfaces. The best conductance was found for
the Si substrate with the thin SiO$_2$ layer. The differences
between the remaining data are mainly due to different wire
heights. In our experiments Si substrates with thin isolation
layers consistently exhibited the lowest fast heating of typically
$\Delta T_f \sim 50$K corresponding to $\sim20$\% resistance
increase at $10^7$A/cm$^2$.

In a two dimensional model for the heat transport within the
substrate, we assume a line-like heat source on the surface of a
half space substrate. The temperature increase $\Delta T_s(t)$ at
this point is then given by the incomplete $\Gamma$ function
\begin{equation}
\Delta T_s(t)=\frac{hw\rho j^2}{2\pi\lambda}\Gamma\left(0,\frac{C
w^2}{4\pi^2\lambda t}\right)\approx \frac{\rho I
j}{2\pi\lambda}\ln\left(\frac{4\pi^2\lambda t}{C
w^2}\right)\label{eq:slow}
\end{equation}
Here, the (small) temperature dependence of the resistivity is
neglected.

For this slow heating process, the total heat dissipation becomes
important. Hence, wider wires heat up faster than narrow ones for
equal current density (Eq.\ \ref{eq:slow} and Fig.\ \ref{crosspic}
B). In addition, the heat transport in Si is faster than in the
other tested materials due to its larger heat conductivity
($\lambda_\mathrm{GaAs} \approx\lambda_\mathrm{sapphire}
\approx\lambda_\mathrm{Si}/3$). Fig.\ \ref{substrpic} shows that
the thin wires ($1.4\mu$m) mounted on Si heat up significantly
less than the taller wires ($\sim 3\mu$m) fabricated on sapphire
or GaAs. In the latter case, the simple model underestimates the
temperature rise for large power dissipation as predicted by the
numerical heat equation integration. The analytical model is only
valid as long as the substrate can be treated as a heat sink. For
a thin substrate (typically 700$\mu$m) the heat transport out of
the substrate has to be taken into account for longer times
(typically after a few 100ms). Then a two-dimensional numerical
calculation accurately reproduces the data (Fig.\ \ref{datapic}).

To conclude, we have presented a method for fabricating atom chips
with a lithographic lift-off process. With this process we
produced wires that can tolerate high current densities of
$>10^7$A/cm$^2$. We have shown that the temperature evolution of
surface mounted microwires agrees with a simple dissipation model.
The optimal substrate has a large heat conductivity and capacity
and is in good thermal contact with the wire. Si substrates with
thin oxide layers showed the best thermal properties of the
examined samples as well as good surface qualities. The described
fabrication process leads to very accurate edge and bulk features.
As a result, the disorder potentials have been observed to be
sufficiently small not to fragment a cold thermal atomic cloud
($T=1\mu$K) at a distance of $<5\mu$m from the wire \cite{Kru04}.

We thank T. Maier and K. Unterrainer of the Mikrostrukturlabor der
TU-Wien for valuable discussions and their insight in developing
the initial fabrication process this work is based on, O. Raslin
for her help in the fabrication process and C. Becker and A. Mair
for their help with the test measurements. This work was supported
by the European Union, contract numbers IST-2001-38863 (ACQP) and
HPRI-CT-1999-00114 (LSF) and the Deutsche Forschungsgemeinschaft,
Schwerpunktprogramm `Quanteninformationsverarbeitung'.



\end{document}